%% file: main.tex
\documentclass[conference]{IEEEtran}
\usepackage{mmap}
\usepackage{soul}
\IEEEoverridecommandlockouts

\usepackage[utf8]{inputenc}

\usepackage{cite}
\usepackage{amsmath,amssymb,amsfonts}
\usepackage{graphicx}
\usepackage{textcomp}
\usepackage{verbatim}
\usepackage{textcomp}
\usepackage{multirow}
\usepackage{pbox}
\usepackage{subcaption}
\usepackage{color}
\usepackage{url}
\usepackage{adjustbox}

\usepackage{listings}
\usepackage{amssymb}
\usepackage{pifont}
\newcommand{\cmark}{\ding{51}}%
\newcommand{\xmark}{\ding{55}}%

\usepackage[ruled,lined,vlined,linesnumbered]{algorithm2e}

\def\BibTeX{{\rm B\kern-.05em{\sc i\kern-.025em b}\kern-.08em
    T\kern-.1667em\lower.7ex\hbox{E}\kern-.125emX}}

\author{
\IEEEauthorblockN{Roman Vasiliev\IEEEauthorrefmark{1}, Dmitrij Koznov\IEEEauthorrefmark{2}, George Chernishev\IEEEauthorrefmark{2}, Aleksandr Khvorov\IEEEauthorrefmark{1}\IEEEauthorrefmark{3}, Dmitry Luciv\IEEEauthorrefmark{2}, Nikita Povarov\IEEEauthorrefmark{1}}
\IEEEauthorblockA{\IEEEauthorrefmark{1}\textit{JetBrains, Saint-Petersburg, Russia}\\
\texttt{\{roman.vasiliev, aleksandr.khvorov, nikita.povarov\}@jetbrains.com}}
\IEEEauthorblockA{\IEEEauthorrefmark{2}\textit{Saint Petersburg State University, Russia}\\
\texttt{\{d.koznov, g.chernyshev, d.lutsiv\}@spbu.ru}}
\IEEEauthorblockA{\IEEEauthorrefmark{3}\textit{ITMO University, Russia}}
}
\title{TraceSim: A Method for Calculating Stack~Trace~Similarity}

\begin{document}
\maketitle
\begin{abstract}
Many contemporary software products have subsystems for automatic crash reporting. However, it is well-known that the same bug can produce slightly different reports.  To manage this problem, reports are usually grouped, often manually by developers. Manual triaging, however, becomes infeasible for products that have large userbases, which is the reason for many different approaches to automating this task. 
Moreover, it is important to improve quality of triaging  due to the big volume of reports that needs to be processed properly. Therefore,  even a relatively small improvement could play a significant role in overall accuracy of report bucketing.
The majority of existing studies use some kind of a stack trace similarity metric, either based on information retrieval techniques or string matching methods. However, it should be stressed that the quality of triaging is still insufficient. 

In this paper, we describe TraceSim~--- a novel approach to address this problem which combines TF-IDF, Levenshtein distance, and machine learning to construct a similarity metric. Our metric has been implemented inside an industrial-grade report triaging system. The evaluation on a manually labeled dataset shows significantly better results compared to baseline approaches. 





\end{abstract}


\section{Introduction}

\input{introduction}

\section{Background}

\input{background}

\section{Related work}

\input{related}

\section{Algorithm}\label{sec:simcalcalg}

\input{algorithm}

\section{Evaluation}

\input{evaluation}

\section{Conclusion}

\input{conclusion}

\bibliographystyle{plain}
\bibliography{bibliography}

\end{document}

%% file: introduction.tex
Systems for collecting and processing bug feedback are nearly ubiquitous in software development companies. However, writing bug reports may require substantial effort from users. Therefore, in order to reduce this effort, a way to create such reports automatically is implemented in most widely used products. In most cases, information available at the time of the crash, i.e. stack trace, is used to form a report.

The drawback of this approach is the huge number of generated reports, the majority of which are duplicates. For example, the study~\cite{microsoft} describes WER~--- the system used in Microsoft to manage crash reports. This system had collected billions of reports from 1999 to 2009. Another example is the Mozilla Firefox browser: according to the study~\cite{Campbell:2016:UET:2901739.2901766}, in 2016 Firefox was receiving $2.2$ million crash reports a day.

It was demonstrated~\cite{dhaliwal} that correct automatic assignment has a positive impact on the bug fixing process. Bugs whose reports were correctly assigned to a single bucket are fixed quicker, and, on the other hand, bugs with reports that were ``spread'' over several buckets take a longer time to fix. 

Thus, the problem of automatic handling of duplicate crash reports is relevant for both academia and industry. There is already a large body of work in this research area, and providing its summary can not be easy, since different studies employ different problem formulations. However, the two most popular tasks concerning automatically created bug reports are:
\begin{enumerate}
	\item for a given report, find similar reports in a database and rank them by the likelihood of belonging to the same bug (ranked report retrieval)~\cite{modani, brodie};
	\item distribute a given set of reports into buckets (report clusterization)~\cite{7866753}.
\end{enumerate}

For both of these tasks, defining a good similarity measure is a must, since the quality of the output largely depends on it. Moreover, it is important to improve similarity algorithms carefully due to the big volume of reports that needs to be processed properly. Even a relatively small improvement could play a significant role in increasing the quality of report bucketing.

In this paper, we address the problem of computing the similarity of two stack traces. The majority of deduplication studies can be classified into two groups: based either on TF-IDF or stack trace structure. The former use an information retrieval approach, while the latter employ string matching algorithms (such as edit distance) to compute stack trace similarity. However, to the best of our knowledge, there are no studies that offered a proper, non-naive combination of these two approaches. Such combination may result in a superior quality of bucketing and may significantly outperform any method that belongs to these individual groups. To substantiate importance of this idea, we would like to quote Campbell et al.~\cite{Campbell:2016:UET:2901739.2901766}: ``a technique based on TF-IDF that also incorporates information about the order of frames on the stack would likely outperform many of the presented methods...''. 

At the same time, machine learning (ML) was rarely applied to this domain: the majority of existing similarity calculation methods does not rely on ML techniques. The reason behind this is the fact that classic (non-ML) methods are more robust and stable than ML ones, which is very important for the considered task. Therefore, our idea is to use classic approaches as the basis.

However, ML methods are more flexible and their application allowed to achieve substantial results in many areas of software engineering. Here, in this particular problem, moderately employing ML allows us to efficiently integrate both classic approaches. Therefore we believe that combining all three approaches would allow us to design superior similarity function.






The contribution of this paper is TraceSim~--- the first algorithm for computing stack trace similarity that structurally combines TF-IDF~\cite{sparck1972statistical}, and string distance while using machine learning to improve quality.


We validate our algorithm using a real-life database of crash reports collected for JetBrains products.

%% file: background.tex
\label{sec:stacktrace}

\subsection{Stack traces}

\begin{figure}[h!]
\includegraphics[width=87mm,height=5cm]{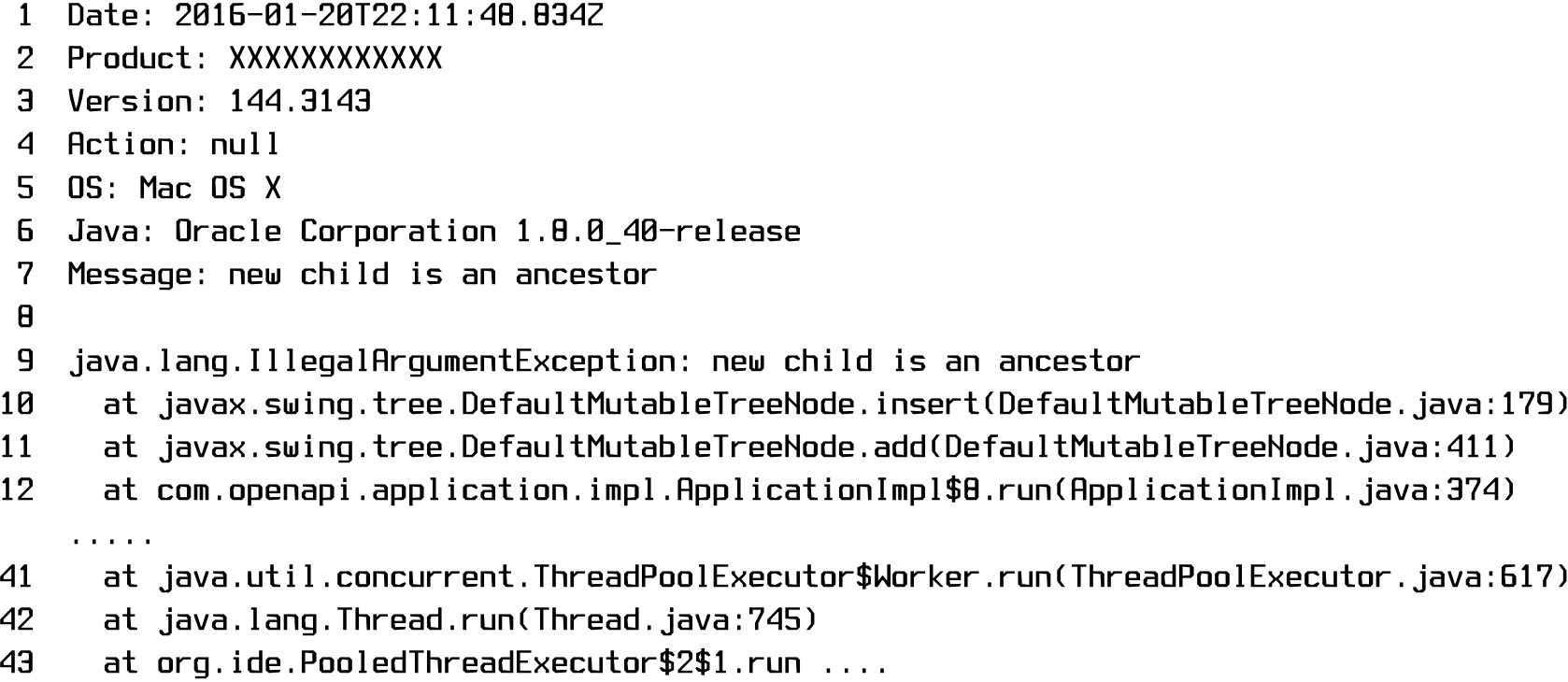}
\caption{Crash report example}\label{fig:crashreport}
\end{figure}

When a contemporary application crashes, it generates a crash report with the following information: application details, environment details, and crash location. In this paper, we are going to examine Java exceptions only.  Crash location information is represented by a stack trace~--- a snapshot of the application call stack that was active at the time of the crash. For each entry of a call stack, its qualifier and line number where a function was called or an error was raised are recorded and stored in the stack trace. The first frame of the stack trace corresponds to the top of the call stack, i.e. to the exact method where the error was raised. Next, there is a sequence of frames which correspond to other methods from the call stack. These go up to the ``main'' function or thread entry function. We will denote a stack trace that contains $N$ frames as $ST = f_0,\ldots,f_{N-1}$. An example of a crash report is presented in Fig.~\ref{fig:crashreport}.

Here, \textbf{lines~1--7} contain the general crash information: the version and the build number of the product, as well as the versions of the operating system and Java. ~\textbf{Line~9} describes the type of the exception and contains the message. Next, the stack trace \(ST = f_0,\ldots,f_{33}\) for the given crash record is presented in~(\textbf{lines~10--43}). 

\subsection{Crash Reports, Stack Traces and Software Quality}

Currently, systems that automatically collect crash stack traces from remote instances are very popular for mass-deployed applications. Prominent examples of such systems are Mozilla Socorro\footnote{\url{https://crash-stats.mozilla.com/}}, LibreOffice Crash Reports\footnote{ \url{http://crashreport.libreoffice.org/}}, Google Chromium\footnote{https://goto.google.com/crash/root} crash reporting system, Windows Error Reporting~\cite{microsoft, dang} and many others. These systems are not a substitute to traditional bug trackers, but are an addition. They are tightly integrated with bug trackers in order to link stack traces to existing bugs, to form new bugs out of a collection of stack traces, and so on.

Having this kind of system allows to obtain bug feedback without requiring users to form and submit ``classic'' bug reports. This, in turn, reduces the strain put on users and allows to greatly increase the amount of collected feedback, which is used to improve software development process. Overall, the benefits are the following:

\begin{itemize}
    \item It allows to survey bug landscape at large at any given moment. For example, LibreOffice Crash Reports show\footnote{https://crashreport.libreoffice.org/stats/} the aggregated view of all received reports over the last $N$ days.
    \item It helps to locate bug in the source code. Both Mozilla Socorro and LibreOffice Crash Reports are integrated with projects' repositories. A user can click on stack frames that are attached to a bug and be transferred to the corresponding lines in the source code.
    \item It allows to automate bug to developer assignment. For example, ClusterFuzz\footnote{https://google.github.io/clusterfuzz/} system allows to automatically assign bug to developer based on crash location in the source code.
\end{itemize}

Crash report management is, therefore, deeply incorporated in the contemporary product development workflow.

All the above mentioned use-cases require to manage harvested stack traces which includes collecting, storing, and retrieving. In its turn, for all these operations to be efficient it is necessary to be able to compare stack traces with respect to bugs that spawn them. 

The challenge is not only the large number of reports, but also the ubiquitous presence of exact and more
importantly, inexact duplicates. For example, our internal study found that 72\% of crash reports of the IntelliJ Platform (a JetBrains product) are duplicates. Due to a large volume of data it is necessary to have a high-quality stack trace similarity measure in order to eliminate duplicates and to group similar crash reports together. Therefore, such measure has great impact on ensuring quality of the software product.


%% file: related.tex
\label{sec:related}

In our survey, we restrict ourselves to reviewing studies that present systems using an explicit similarity function for bucketing reports based on their stack traces. Surveys on triaging involving textual descriptions and tags can be found in~\cite{Hindle2018,8048025,6100061}. 

Since we are interested in constructing a novel stack trace similarity measure that will combine TF-IDF, edit distance, and supervised machine learning approaches we highlight the respective components of existing studies. The big picture is presented in Table~\ref{tab:relwork}.

A study by Brodie et al.~\cite{brodie} was one of the earliest that addressed the problem of crash report deduplication using stack trace comparison. They present a biological sequence search algorithm that is a modification of the Needleman-Wunsch algorithm~\cite{NEEDLEMAN1970443}.


Bartz et al.~\cite{bartz} construct a callstack similarity measure that is essentially a modification of the edit distance metric, and propose seven edit operations with different weights. 

Dhaliwal et al.~\cite{dhaliwal} propose a two-step approach that combines signatures and the Levenshtein distance between stack traces. The idea is the following: first, reports are grouped together using only the first frame. Then, each bucket is split into several subgroups using the Levenshtein distance between stack traces (only the top 10 frames are used). 

Kim et al.~\cite{5958261} use stack traces contained in a bucket, building a special graph on the base of similarity of two stack traces, and then applying a graph similarity measure to decide whether the new stack trace belongs to this bucket. 

Modani et al.~\cite{modani} compare three methods for calculating stack trace similarity: edit distance, prefix match and Longest Common Subsequence (LCS). 

Dang et al.~\cite{dang} present a new similarity metric that is based on the offset distance between the matched functions and the distance from these functions to the top frame. This method employs edit distance and relies on supervised learning approach.

Lerch and Mezini~\cite{Lerch:2013:FDY:2495256.2495763} study the situation when a bug tracker does not contain a dedicated field for storing crash stacks. The authors employ the TF-IDF approach for finding duplicate stack traces. 

Wu et al.~\cite{Wu:2014:CLC:2610384.2610386} adapt the TF-IDF approach by introducing the notions of function frequency and inverse bucket frequency. A notable idea of this approach is to expand the list of functions present in the stack trace by adding the ones that are likely to be the root cause of the crash. For this, a technique comprised of control flow analysis, backward slicing, and function change information is proposed.

Campbell et al.~\cite{Campbell:2016:UET:2901739.2901766} compare automatic crash report deduplication methods. The authors have considered two types of algorithms: TF-IDF-based (using ElasticSearch ~\cite{Elasticsearch}) and signature-based. The results of the evaluation demonstrate the superiority of information retrieval methods.

Moroo et al.~\cite{DBLP:conf/seke/MorooAH17} propose a reranking-based crash report clustering method. It is a combination of two state-of-the-art report deduplication methods: ReBucket~\cite{dang} and Party-Crasher~\cite{Campbell:2016:UET:2901739.2901766}. Since this method employs ReBucket as its part, it has both edit distance and supervised learning components. Party-Crasher part supplies TF-IDF element. However, authors propose a straightforward technique which essentially invokes these two approaches independently and then computes their a weighted harmonic mean. While experiments demonstrated that such technique can be superior to its constituent parts, it is still not a proper structural integration. It is possible that an algorithm with ``true'' structural integration of TF-IDF and edit distance components (i.e. that describes more sophisticated relation between them) may yield significantly better results.


Finally, Sabor et al.~\cite{8009929} proposed DURFEX system which combined stack trace similarity and a similarity of two non-textual fields. For computing stack trace similarity authors proposed to substitute function names by names of packages where they are defined and then to segment the resulting stack traces into N-grams of variable length.

We can see that despite the fact that this problem has been studied for at least 15 years, it is still relevant for the community: new studies continue to emerge. Although there is a variety of methods, the quality of bucketing continues to be insufficient. Finally, we can see that there are no approaches that perform structural integration of edit distance and TF-IDF, despite this combination looking promising~\cite{Campbell:2016:UET:2901739.2901766} in a sense that it may substantially improve the quality of bucketing.

\begin{table}
\caption{Existing approaches}\label{tab:relwork} 
\begin{center}
     \begin{tabular}{| l | c | c | c |}
          \hline
           \textbf{Method} & \textbf{TF-IDF} & \textbf{Edit Distance} & \textbf{Machine Learning} \\ \hline
           
           Brodie et al.~\cite{brodie} & \xmark & \cmark & \xmark \\ \hline
           
           Bartz et al.~\cite{bartz} & \xmark & \cmark & \cmark \\ \hline
           
           Dhaliwal et al.~\cite{dhaliwal} & \xmark & \cmark & \xmark \\ \hline
           
           Kim et al.~\cite{5958261} & \xmark & \xmark & \cmark \\ \hline
           
           Modani et al.~\cite{modani} & \xmark & \cmark & \xmark \\ \hline
           
           Dang et al.~\cite{dang} & \xmark & \cmark & \cmark \\ \hline
           
           Lerch and Mezini~\cite{Lerch:2013:FDY:2495256.2495763} & \cmark & \xmark & \xmark \\ \hline
           
           Wu et al.~\cite{Wu:2014:CLC:2610384.2610386} & \cmark & \xmark & \xmark \\ \hline

           Campbell et al.~\cite{Campbell:2016:UET:2901739.2901766} & \cmark & \xmark & \xmark \\ \hline

           Moroo et al.~\cite{DBLP:conf/seke/MorooAH17} & \cmark & \cmark & \cmark \\ \hline
           
           Sabor et al.~\cite{8009929} & \xmark & \xmark & \cmark \\ \hline
           
      \end{tabular}
    \end{center}
 \end{table}

%% file: algorithm.tex
\label{sec:algo}

\newcommand{\lwp}{\alpha}
\newcommand{\gwp}{\beta}
\newcommand{\threshold}{\gamma}

\newcommand{\lwa}{\operatorname{\mathbf{lw}_{\lwp}}}
\newcommand{\gw}{\operatorname{gw}}
\newcommand{\gwbc}{\operatorname{\mathbf{gw}_{\gwp\threshold}}}
\newcommand{\sff}{\operatorname{sff_{bc}}} 

In this section, we describe our algorithm for computing stack trace similarity. Our algorithm takes two stack traces as its input. First, it processes stack overflow exceptions (SOEs) separately, since these stack traces contain a large number of repeated frames, and their similarity can be calculated effectively using TF-IDF (here we used approach from ~\cite{Lerch:2013:FDY:2495256.2495763}). If the input stack traces are not SOEs, the algorithm proceeds to compute their similarity in a different way. First, it computes the weight for each frame of the stack traces, because different frames have different impacts on stack trace similarity. Next, the edit distance between two stack traces is calculated. In our approach, this distance is defined as Levenshtein distance~\cite{levenshtein} with frame weights.
Finally, the results are normalized using the calculated Levenshtein distance. An implementation of the TraceSim algorithm can be found here~\cite{TraceSim}. 

A detailed description of the above steps follows.

\subsection{Separate processing of SOEs}\label{sub:SOEs}

A stack trace that is a stack overflow exception contains many repeated frames which refer to recursive calls. If this recursive part of two stack traces is similar, it is highly probable that they address the same error situation. Usually, this recursive part is rather large, significantly exceeding the non-recursive part of the stack trace in size. Therefore, complicated tests are unnecessary for such stack traces, and computing their closeness in terms of frame frequencies is enough. This is the reason we use the TF-IDF algorithm from~\cite{Lerch:2013:FDY:2495256.2495763} in this case. 

\subsection{Frame weight computation}\label{sub:similarity}

While comparing two stack traces, differences in frames that are close to the top of the stack are usually more important than differences in deeper-positioned frames~\cite{schroter}. We propose to represent this influence as frame weight: frames with higher weights are considered more important. We identify two factors that affect frame weight: frame position within a stack trace and frame frequency among all frames of all stack traces available in our database. For a stack frame \(f_i\) of \(ST=f_0,\ldots,f_{N-1}\), its weight is calculated as follows:

\begin{equation}
\operatorname{\mathbf{w}}(f_i) = \lwa(f_i)*\gwbc(f_i),
\end{equation}

\noindent{}where \(\lwa(f_i)\) is the local weight of \(f_i\), i.e. the degree of its importance among other frames of the same stack trace, and \(\gwbc(f_i)\) is the global weight of the frame, i.e. the degree of its importance among all frames of all stack traces presenting in our database. Here, $\lwp$, $\gwp$ and $\threshold$ are numeric hyperparameters~\cite{claesen2015hyperparameter}.



Local frame weight of \(f_i\) is calculated as follows:

\begin{equation}
\lwa(f_i)= \frac{1}{i^\lwp}
\end{equation}

Local weight is higher for frames which are closer to the top of the stack, since as practice shows, these frames are more important than further ones, i.e., errors are more likely caused by the functions which were called last.

Global frame weight of \(f_i\) is calculated according to the well-known information retrieval TF-IDF approach~\cite{Manning:2008:IIR:1394399} as \(\operatorname{TF}(f_i) * \operatorname{IDF}(f_i)\), where \(\operatorname{TF}(f)\) (term frequency) represents the importance of the frame within a particular stack trace, while \(\operatorname{IDF}(f)\) (inverse document frequency) represents how uncommon is the frame \(f\) for the whole corpus of stack traces.
In our work, we do not use the \(\operatorname{TF}\) part and consider it equal to~1 since it does not consider frame ordering, which is actually the most important information about the frame within the stack trace. This has already been taken into account when calculating \(\lwa(f_i)\). Hence, we only calculate \(\operatorname{IDF}(f_i)\) as

\begin{equation*}
\operatorname{IDF}(f_i) =
\log\frac{\text{Total num. of stack traces}}{\text{Num. of stack traces}\; ST: f_i \in ST}.
\end{equation*}

Therefore, we calculate global weight as follows:

\begin{equation}
\gwbc(f_i) = \operatorname{sigm}(\gwp(\operatorname{IDF}(f_i) - \threshold)),
\end{equation}\label{eq:gr}

where $\operatorname{sigm}$ is a sigmoid function defined as:

\begin{equation}
\operatorname{sigm}(x) = \frac{1}{1 + e^{-x}}.
\end{equation}\label{eq:gr}

Here, the \(\gwp\) and \(\threshold\) hyperparameters are used to tune smooth filtering for $\operatorname{IDF}(f_i)$. We give small weights for very common frames that are contained in a large number of stack traces. Those can be frames that emerge due to frequently invoked chunks of code: commonly used development frameworks, logging or thread pooling.


\subsection{Levenshtein distance calculation}

In order to express difference between stack traces numerically, we use modified Levenshtein distance. As the basis we took classic Levenshtein distance that contains only insertion, deletion, and replacement operators. We do not consider a variation that includes transposition operation, since that for stack traces the order of the frames is very important: swapping places of two frames within a single stack trace is meaningless.

For two strings, classic Levenshtein distance is defined as minimal editing cost, i.e. the minimal total number of insertions, deletions, and replacements of a single character needed to transform one string into another~\cite{levenshtein}. For two stack traces, we define the distance in the same way, but additionally using the weights assigned to frames: stack traces that differ in ``heavy'' frames are more different themselves.

When calculating the cost of insertion, deletion or substitution of a frame, we define operation costs as follows: cost of insertion and deletion is the weight of the corresponding frame and the weight of substitution is the sum of weights of the original and the new frame. 
\subsection{Normalization}

We do not use the Levenshtein distance itself for classification and clustering, instead, we calculate a normalized similarity value:

\begin{equation}
\operatorname{sim}(ST', ST'') = 1 -
\frac{\operatorname{dist}(ST', ST'')}%
{\sum_{i=0}^{N'-1}\operatorname{w}(f'_i) + \sum_{i=0}^{N''-1}\operatorname{w}(f''_i)},
\end{equation}

\noindent{}where \(\operatorname{dist}(ST', ST'')\) stands for the Levenshtein distance between \(ST'=f'_0,\ldots,f'_{N'-1}\) and \(ST''=f''_0,\ldots,f''_{N''-1}\).

\subsection{Hyperparameter Estimation via Machine Learning}

In previous subsections we have introduced $\lwp$, $\gwp$ and $\threshold$~--- numeric hyperparameters used in calculation of local and global frame weights. To obtain their values we formulate an optimization problem and approach it with machine learning. The idea is to optimize ROC AUC~\cite{Fawcett2006} metric by training on a manually labelled part of the stack trace dataset. To solve it, the Tree-structured Parzen Estimator Approach (TPE)~\cite{10.5555/2986459.2986743} was used, we have employed  hyperopt\footnote{\url{https://github.com/hyperopt/hyperopt}}~\cite{10.5555/3042817.3042832} library.

%% file: evaluation.tex
\label{sec:eval}

\subsection{Experimental Setup}

To perform the evaluation, we have used the JetBrains crash report processing system Exception Analyzer which handles reports from various IntelliJ Platform products. Exception Analyzer receives generated reports and automatically distributes them into existing issues (buckets) or creates new ones out of them. However, it is a well-known problem that output of automatic bug triaging tools can be of insufficient quality~\cite{dang}. Therefore, Exception Analyzer allows to employ user input to triage ``problematic'' reports. If this happens, user actions are logged and can be used later on for various purposes.


In order to evaluate our approach, we had to construct a test corpus. We adhere to the following idea: if a developer assigns a report to an issue manually, then there is a reason to think that this report is a duplicate to the ones already contained in the issue. And vice versa: if a developer extracts some reports from a given issue, it means that these reports are distinct from the remaining.

To construct our test corpus, we have extracted and analyzed reports from recent user action logs of Exception Analyzer spanning one year time frame. To create positive pairs we have analyzed user sessions and searched for the following pattern: for a particular unbucketed report, a user looks into some issue, compares it to a particular report of this issue and then assigns it into the issue. To obtain negative pairs we exploit a similar idea: we designate a pair as negative if a user compared reports and did not grouped them. Eventually, we have obtained 6431 pairs, out of which 3087 were positive and were 3344 negative. We have got not too many pairs due to the fact that most reports in  Exception Analyzer are grouped automatically and users rarely have to intervene. The experiments were run with 80/20 test-train split.

\subsection{Research Questions}

\begin{itemize}{\setlength\itemindent{10pt}
	\item[\textbf{RQ1:}] How do individual steps contribute to the overall quality of algorithm output?
	\item[\textbf{RQ2:}] How well does our approach perform in comparison with state-of-the-art approaches?
}\end{itemize}

RQ1 evaluates the effectiveness of individual components of our method. Since our function consists of a number of independent steps, it is necessary to check whether each of them is beneficial or not. By performing these evaluations, we demonstrate that each component is essential for our resulting similarity function.
We perform several experiments for this purpose. For every experiment, we switch off the corresponding component in the full TraceSim, and run it on the test corpus. We consider the following steps: TraceSim without gw ($\operatorname{\mathbf{gw}}(f_i)) = 1$ in (1)), TraceSim without lw ($\lwa(f_i)=1$ in (1)), TraceSim without SOEs (without separate processing of stack overflow exceptions using the algorithm from~\cite{Lerch:2013:FDY:2495256.2495763}), and the full version of TraceSim.


RQ2 compares the resulting similarity function with the state-of-the-art approaches. 

First, we considered approaches that use TF-IDF~\cite{Lerch:2013:FDY:2495256.2495763, DBLP:conf/seke/MorooAH17} technique. Next, we also employed Rebucket~\cite{dang} method and its available implementation~\cite{RebucketImplementation}. It should be noted that it belongs to edit distance and supervised learning methods. We also included in our baseline other edit distance methods~--- Levenshtein distance~\cite{modani} and Brodie~\cite{brodie}. Another supervised method that we included in our evaluation is Moroo et al.~\cite{DBLP:conf/seke/MorooAH17}. It combines Rebucket and Lerch et al.~\cite{Lerch:2013:FDY:2495256.2495763} approaches.

We didn't compared with recently-developed DURFEX~\cite{8009929} approach since it relies on tight integration with bug tracker and requires component and severity fields. At the same time our approach concerns only stack traces.

Finally, we have decided to compare our approach with several classic and widely known approaches: Prefix Match and Cosine Similarity~\cite{modani}. We have employed two variations of the latter: Cosine Similarity with IDF component (denoted as Cosine (IDF)) and without (denoted as Cosine (1)).

\subsection{Evaluation Metrics}

To answer RQs 1 and 2, we have evaluated how good our similarity function is. Due to nature of our dataset we have to use metric applicable for binary classification. To assess the quality of our algorithm, we use the well-accepted comparison measure ROC AUC~\cite{Ling:2003:ASC:1630659.1630736}. It is statistically consistent, and it is also a more  discriminating measure than Precision/Recall, F-measure, and Accuracy. Several studies concerning bug report triage also employ metrics like MAP~\cite{10.1007/s10664-018-9643-4}, Recall Rate~\cite{8094414, 6062072}, and other metrics used for the ranking problem. However, in this paper we consider the binary classification task and therefore we need to use other metrics.

Turning to ROC AUC, an important observation that 0.5 is considered the minimum result for ROC AUC due to simple random classifier giving a result of 0.5. In our experiments we didn't used cross validation since we have sufficient data to run a simple test/train split.


Another observation is the following: if an algorithm increases ROC AUC from $0.5$ to $0.55$, this increase is less significant than the one from $0.75$ to $0.8$, despite the equal gain. This is the reason why we have computed the error reduction of each algorithm (RQ2). After ranging the algorithm outputs by ROC AUC, we calculate by how many percent the error rate has been reduced in comparison to the previous algorithm. For example, method of Brodie et al. has improved by $0.06$ in comparison to Prefix Match (0,64 against 0,58), and its error reduction is $0.06*100/(1-0.58)=14\%$. These numbers are presented in Table~\ref{tab:comparison}.

\subsection{Results}

\subsubsection{RQ1: How do individual steps contribute to the overall quality of the algorithm output?}


The ROC AUC results are presented in Table~\ref{tab:steps}. We have found out that the $\operatorname{\mathbf{gw}}$ weight function, which is based on computing global frequency for frames, makes the largest contribution ($+0.1$). The $\operatorname{\mathbf{lw}}$  weight function that considers the order of frames in a stack trace contributes less ($+0.03$). Finally, SOEs contribute the least ($+0.01$), which is explicitly connected to the number of stack traces containing recursion ($4\%$ in our test corpus).

\begin{table}
	\caption{Contribution of individual steps}\label{tab:steps}
	\begin{center}	
		\begin{tabular}{| l | l |}
			\hline
			\textbf{Method}  & \textbf{Results}  \\ \hline
			TraceSim & $0.79$ \\ \hline
			TraceSim without SOEs  & $0.78$ \\ \hline
            TraceSim without lw & $0.76$ \\ \hline
			TraceSim without gw & $0.69$ \\ \hline
		\end{tabular}
	\end{center}
\end{table}

\subsubsection{RQ2: How well does our approach perform in comparison to state-of-the-art approaches?}

The ROC AUC results are presented in Table~\ref{tab:comparison}. Our method turned out to be superior to all others.  Our contribution is significant: we have improved by $+0.03$ compared to the existing algorithm with the best result on our dataset. However, it should be noted that the improvement of almost all other algorithms lies between $+0.003$ and $+0.06$. Furthermore, our algorithm provides error reduction of 13\%, and only Brodie et al. provides more. 

\begin{table}
\caption{Comparison with other approaches}\label{tab:comparison} 
\begin{center}
     \begin{tabular}{| l | l | l|}
          \hline
           \textbf{Similarity} & \textbf{ROC AUC} & \textbf{Error red.} \\ \hline
            TraceSim & $0.79$ &  $13\%$ \\ \hline
            Moroo et al.~\cite{DBLP:conf/seke/MorooAH17} & $0.76$ &  $0\%$ \\ \hline
            Lerch~\cite{Lerch:2013:FDY:2495256.2495763} & $0.76$ &  $11\%$ \\ \hline
            Cosine (IDF) & $0.73$ & $10\%$\\ \hline
	        Rebucket~\cite{dang} & $0.70$ & $6\%$\\ \hline
            Cosine (1) & $0.68$ & $0\%$\\ \hline
            Levenshtein~\cite{modani} & $0.68$ & $11\%$\\ \hline
            Brodie et al.~\cite{brodie} & $0.64$ & $14\%$\\ \hline
            Prefix Match~\cite{modani} & $0.58$ & $-$ \\ \hline 

      \end{tabular}
    \end{center}
 \end{table}


%% file: conclusion.tex
\label{sec:conclusion}

In this paper, we have proposed a novel approach to calculating stack trace similarity that combines TF-IDF and Levenshtein distance. The former is used to ``demote'' frequently encountered frames via an IDF analogue for stack frames, while the latter allows to account for differences not only in individual frames, but also in their depth. At the same time, employed machine learning allowed us to efficiently combine two classic approaches.

To evaluate our approach, we have implemented it inside an industrial-grade report triaging system used by JetBrains. The approach has been employed for over 6 months, receiving positive feedback from developers and managers, who reported that the quality of bucketing had improved. Our experiments have shown that our method outperforms the existing approaches. It should be noted that even a relatively small improvement plays a significant role in the quality of report bucketing due to the large overall report volume.
  

%% file: main.bbl
\begin{thebibliography}{10}

\bibitem{Elasticsearch}
Elasticsearch.
\newblock \url{https://www.elastic.co/products/elasticsearch}.
\newblock Accessed: {2019-11-08}.

\bibitem{RebucketImplementation}
Implementation of {R}ebucket.
\newblock \url{https://github.com/ZhangShurong/rebucket}.
\newblock Accessed: {2019-11-08}.

\bibitem{TraceSim}
Trace{S}im implementation.
\newblock \url{https://github.com/traceSimSubmission/trace-sim}.
\newblock Accessed: {2019-11-08}.

\bibitem{bartz}
K.~Bartz, J.~W. Stokes, J.~C. Platt, R.~Kivett, D.~Grant, S.~Calinoiu, and
  G.~Loihle.
\newblock Finding similar failures using callstack similarity.
\newblock SysML'08, pages 1--6, Berkeley, CA, USA, 2008. USENIX Association.

\bibitem{10.5555/2986459.2986743}
J.~Bergstra, R.~Bardenet, Y.~Bengio, and B.~K\'{e}gl.
\newblock Algorithms for hyper-parameter optimization.
\newblock NIPS’11, page 2546–2554, 2011.

\bibitem{10.5555/3042817.3042832}
J.~Bergstra, D.~Yamins, and D.~D. Cox.
\newblock Making a science of model search: Hyperparameter optimization in
  hundreds of dimensions for vision architectures.
\newblock ICML’13, page I–115–I–123. JMLR.org, 2013.

\bibitem{brodie}
M.~Brodie, Sheng Ma, G.~Lohman, L.~Mignet, N.~Modani, M.~Wilding, J.~Champlin,
  and P.~Sohn.
\newblock Quickly finding known software problems via automated symptom
  matching.
\newblock ICAC'05, pages 101--110, June 2005.

\bibitem{Campbell:2016:UET:2901739.2901766}
J.~C. Campbell, E.~A. Santos, and A.~Hindle.
\newblock The unreasonable effectiveness of traditional information retrieval
  in crash report deduplication.
\newblock MSR '16, pages 269--280, New York, NY, USA, 2016. ACM.

\bibitem{claesen2015hyperparameter}
Marc Claesen and Bart De~Moor.
\newblock Hyperparameter search in machine learning.
\newblock {\em arXiv preprint arXiv:1502.02127}, 2015.

\bibitem{dang}
Y.~Dang, R.~Wu, H.~Zhang, D.~Zhang, and P.~Nobel.
\newblock Rebucket: A method for clustering duplicate crash reports based on
  call stack similarity.
\newblock ICSE '12, pages 1084--1093, Piscataway, NJ, USA, 2012. IEEE Press.

\bibitem{8094414}
J.~{Deshmukh}, K.~M. {Annervaz}, S.~{Podder}, S.~{Sengupta}, and N.~{Dubash}.
\newblock Towards accurate duplicate bug retrieval using deep learning
  techniques.
\newblock ICSME '17, pages 115--124, 2017.

\bibitem{dhaliwal}
T.~Dhaliwal, F.~Khomh, and Y.~Zou.
\newblock Classifying field crash reports for fixing bugs: A case study of
  mozilla firefox.
\newblock ICSM '11, pages 333--342, Washington, DC, USA, 2011. IEEE Computer
  Society.

\bibitem{Fawcett2006}
Tom Fawcett.
\newblock An introduction to {ROC} analysis.
\newblock {\em Pattern Recognition Letters}, 27(8):861--874, jun 2006.

\bibitem{7866753}
M.~A. {Ghafoor} and J.~H. {Siddiqui}.
\newblock Cross platform bug correlation using stack traces.
\newblock FIT '16, pages 199--204, Dec 2016.

\bibitem{microsoft}
K.~Glerum, K~Kinshumann, S.~Greenberg, G.~Aul, V.~Orgovan, G.~Nichols,
  D.~Grant, G.~Loihle, and G.~Hunt.
\newblock Debugging in the (very) large: Ten years of implementation and
  experience.
\newblock SOSP '09, pages 103--116, New York, NY, USA, 2009. ACM.

\bibitem{Hindle2018}
Abram Hindle and Curtis Onuczko.
\newblock Preventing duplicate bug reports by continuously querying bug
  reports.
\newblock {\em Empirical Software Engineering}, Aug 2018.

\bibitem{10.1007/s10664-018-9643-4}
Abram Hindle and Curtis Onuczko.
\newblock Preventing duplicate bug reports by continuously querying bug
  reports.
\newblock {\em Empirical Softw. Engg.}, 24(2):902–936, April 2019.

\bibitem{5958261}
S.~Kim, T.~Zimmermann, and N.~Nagappan.
\newblock Crash graphs: An aggregated view of multiple crashes to improve crash
  triage.
\newblock DSN '11, pages 486--493, June 2011.

\bibitem{Lerch:2013:FDY:2495256.2495763}
Johannes Lerch and Mira Mezini.
\newblock Finding duplicates of your yet unwritten bug report.
\newblock CSMR '13, pages 69--78. IEEE Comp. Soc., 2013.

\bibitem{levenshtein}
V.~I. Levenshtein.
\newblock Binary codes capable of correcting deletions, insertions and
  reversals.
\newblock {\em Soviet Physics Doklady}, 10:707--710, 1966.

\bibitem{Ling:2003:ASC:1630659.1630736}
C.~X. Ling, J.~Huang, and H.~Zhang.
\newblock {AUC}: A statistically consistent and more discriminating measure
  than accuracy.
\newblock IJCAI'03, pages 519--524.

\bibitem{Manning:2008:IIR:1394399}
C.~D. Manning, P.~Raghavan, and H.~Sch\"{u}tze.
\newblock {\em Introduction to Information Retrieval}.
\newblock Cambridge University Press, 2008.

\bibitem{modani}
N.~Modani, R.~Gupta, G.~Lohman, T.~Syeda-Mahmood, and L.~Mignet.
\newblock Automatically identifying known software problems.
\newblock ICDEW '07, pages 433--441, Washington, DC, USA, 2007. IEEE Computer
  Society.

\bibitem{DBLP:conf/seke/MorooAH17}
A.~Moroo, A.~Aizawa, and T.~Hamamoto.
\newblock Reranking-based crash report deduplication.
\newblock In X.~He, editor, {\em SEKE ’17}, pages 507--510, 2017.

\bibitem{NEEDLEMAN1970443}
S.~B. Needleman and C.~D. Wunsch.
\newblock A general method applicable to the search for similarities in the
  amino acid sequence of two proteins.
\newblock {\em Journal of Molecular Biology}, 48(3):443--453, 1970.

\bibitem{8048025}
M.~S. Rakha, C.~Bezemer, and A.~E. Hassan.
\newblock Revisiting the performance evaluation of automated approaches for the
  retrieval of duplicate issue reports.
\newblock {\em IEEE Trans. on Soft. Eng.}, 44(12):1245--1268, Dec 2018.

\bibitem{8009929}
K.~K. {Sabor}, A.~{Hamou-Lhadj}, and A.~{Larsson}.
\newblock {DURFEX}: A feature extraction technique for efficient detection of
  duplicate bug reports.
\newblock ICSQRS '17, pages 240--250, 2017.

\bibitem{schroter}
A.~Schroter, A.~Schröter, N.~Bettenburg, and R.~Premraj.
\newblock Do stack traces help developers fix bugs?
\newblock MSR '10, pages 118--121, May 2010.

\bibitem{sparck1972statistical}
Karen Sparck~Jones.
\newblock A statistical interpretation of term specificity and its application
  in retrieval.
\newblock {\em Journal of documentation}, 28(1):11--21, 1972.

\bibitem{6100061}
C.~Sun, D.~Lo, S.~Khoo, and J.~Jiang.
\newblock Towards more accurate retrieval of duplicate bug reports.
\newblock ASE '11, pages 253--262, Nov 2011.

\bibitem{6062072}
C.~{Sun}, D.~{Lo}, X.~{Wang}, J.~{Jiang}, and S.~{Khoo}.
\newblock A discriminative model approach for accurate duplicate bug report
  retrieval.
\newblock ICSE '10, pages 45--54, 2010.

\bibitem{Wu:2014:CLC:2610384.2610386}
R.~Wu, H.~Zhang, S.~Cheung, and S.~Kim.
\newblock Crashlocator: Locating crashing faults based on crash stacks.
\newblock ISSTA '14, pages 204--214, 2014.

\end{thebibliography}
